# Large Electronic Anisotropy and Enhanced Chemical Activity of Highly Rippled Phosphorene

Andrey A. Kistanov[†,‡], Yongqing Cai[‡,*], Kun Zhou[†,*], Sergey V. Dmitriev [§,∥], Yong-Wei Zhang[‡]

[†]School of Mechanical and Aerospace Engineering, Nanyang Technological University, 639798 Singapore

[‡]Institute of High Performance Computing, A*STAR, 138632 Singapore

[§]Institute for Metals Superplasticity Problems, Russian Academy of Science, 450001 Ufa, Russia

[∥]National Research Tomsk State University, 634050 Tomsk, Russia

**ABSTRACT:** We investigate the electronic structure and chemical activity of rippled phosphorene induced by large compressive strains, via first-principles calculation. It is found that phosphorene is extraordinarily bendable, enabling the accommodation of ripples with large curvatures. Such highly rippled phosphorene shows a strong anisotropy in electronic properties. For ripples along the armchair direction, the band gap changes from 0.84 to 0.51 eV for the compressive strain up to -20% and further compression shows no significant effect; for ripples along the zigzag direction, semiconductor-to-metal transition occurs. Within the rippled phosphorene, the local electronic properties, such as the modulated band gap and the alignments of frontier orbitals, are found to be highly spatially dependent, which may be used for modulating the injection and confinement of carriers for optical and photovoltaic applications. The examination of the interaction of a physisorbed NO molecule with the rippled phosphorene under different compressive strains shows that the chemical activities of the phosphorene are significantly enhanced at the top and bottom peaks of the ripples, indicated by the enhanced adsorption and charge transfer between them. All these features can be ascribed to the effect of curvatures, which modifies the orbital coupling between atoms at the ripple peaks.

## 1. INTRODUCTION

Recent fabrication of two-dimensional (2D) black phosphorus, called phosphorene, by liquid phase exfoliation[1] and mechanical cleavage[2,3] has triggered great interest for its potential applications in nanoelectronics and photonics. Within the phosphorene layer, atoms are bonded together via strong covalent bonds, while between the layers, atoms interact via a weak van der Waals (vdW) force. Its atomically thin structure, similar to graphene and $MoS_2$ albeit buckled, allows phosphorene to withstand large deformation without breaking[4]. Exciting physical properties, such as a direct band gap, high carrier mobility, and asymmetric electronic and phononic transport[5] make phosphorene promising for applications in nanoelectronics devices, including solar cell[6] and field effect transistors (FET)[7,8].

Many studies have been performed to investigate the effect of various factors on the electronic properties of phosphorene[9-15]. For example, phosphorene can be easily oxidized or affected by environments, which may cause the degradation and disintegration of its lattice structure[16]. To prevent phosphorene from structure degradation, passivation using more stable 2D materials such as graphene or h-BN has been proposed to protect its fragile structure[17]. It is also known that the electronic and optical properties of phosphorene can be modulated by defect and strain engineering[18-22], electric field[23], edge functionalization[24] and adatom adsorption[25,26]. For example, Peng *et al.*[27] showed that in-plane strain causes a direct-indirect band gap transition in phosphorene. In addition, they also observed strongly anisotropic transport of carriers, which can be dramatically tuned by strain. However, most of the studies investigated the in-plane strain effects in phosphorene, the effects of ripples or wrinkles with significant out-of-plane deformation, which is highly possible in 2D materials, are rarely studied. Kou *et al*[19]. examined a shallow ripple deformation in phosphorene induced by up to 10% compressive strain, and found that the ripple deformation is anisotropic, occurring only along the zigzag direction, but not along the armchair direction. We note that the effect of large compressive strain (higher than -10%) is still unexplored and the understanding of the effect of deep ripples on the spatially dependent electronic properties of phosphorene is still lacking. Recent experiment demonstrated that ripples in phosphorene could lead to interesting quantum confinement and funneling of carriers for photovoltaic applications[28]. Hence, the knowledge of local electronic properties in deeply rippled phosphorene is not only of scientific significance and but also of great implication for its potential device applications.

Due to their large surface area, atomically thin 2D materials exhibit a high chemical activity to accommodate foreign atoms and make them a promising candidate for gas sensing[25,29]. Many previous theoretical and experimental studies demonstrated the possibility of using graphene[30-32], silicene[33], germanene[34] and $MoS_2$[35-38] as highly sensitive gas sensors. For phosphorene, however, there are only a few studies focusing on its chemical activities and sensing applications[25,39]. Adsorption of organic molecules on phosphorene under external electric field and in-plane strain was investigated[40]. It was shown that the band gap and the polarity of molecularly absorbed phosphorene can be effectively modulated. Due to their extremely low bending rigidity, atomically thin materials can be easily bended either intentionally or unintentionally to form ripples or curved configurations. As demonstrated by previous studies in graphene and carbon nanotube[41-47], these curved structures show significant changes in the adsorption behavior of external chemical species like molecules and atoms[30,48,49]. However, the effect of ripples or wrinkles on the chemical activity of phosphorene is still unknown, despite their inevitable occurrence in devices and fabrications.

In this work, we systematically address the issues discussed above. First of all, a detailed study of the evolution of the atomic structures and electronic properties of rippled phosphorene under large compressive strains (higher than -20%) is presented. Secondly, the changes in the chemical activity of rippled phosphorene upon the adsorption of NO gas molecule are examined. We find that the rippled structure shows a significant modification in the electronic properties compared with its flat counterpart, and the local electronic properties for the atoms located along the periodic rippled profile, such as the modulated band gap and the alignments of frontier orbitals, are found to be highly spatially dependent. Finally, the rippled structure shows a significantly enhanced chemical activity at top and bottom peaks due to the highly tensile or compressive local strains experienced by the atoms located there. The present study suggests that highly rippled phosphorene with unique electronic and sensing properties are promising for novel device applications.

## 2. COMPUTATIONAL METHODS

The spin-polarized first-principles calculations are performed by using the plane-wave Vienna *ab initio* simulation package (VASP)[50] within the framework of density functional theory (DFT). The exchange-correlation functionals are selected as the Perdew-Burke-Ernzerhof functional (PBE)[51] under the generalized gradient approximation (GGA) and the hybrid functional (HSE06). Dispersive interactions during the noncovalent chemical functionalization of phosphorene with small molecules are analyzed using the vdW corrected functional with Becke88 optimization (optB88)[52]. All the structures are fully relaxed until the force is smaller than 0.01 eV/Å. The relaxed lattice constant of monolayer phosphorene is a = 3.335 Å and b = 4.571 Å along zigzag and armchair directions, respectively, based on a $1 \times 4 \times 1$ and $6 \times 1 \times 1$ grid for *k*-point sampling, respectively. Rippled structures are created by using the $10 \times 1 \times 1$ and $1 \times 10 \times 1$ supercells (40 phosphorus atoms) respectively for applying strain along zigzag and armchair directions. A $10 \times 3 \times 1$ supercell is adopted for single NO adsorption. Periodic boundary conditions are applied in the two in-plane transverse directions, whereas free boundary conditions are applied along the out-of-plane direction by introducing a vacuum space of 20 Å. The first Brillouin zone is sampled with a $1 \times 10 \times 1$ ($10 \times 1 \times 1$) *k*-mesh grid for the $10 \times 1 \times 1$ ($1 \times 10 \times 1$) supercell and $1 \times 3 \times 1$ grid for the $10 \times 3 \times 1$ supercell. Kinetic energy cutoff of 450 eV is adopted. The absorption energy ($E_a$) of a molecule on rippled phosphorene is calculated as $E_a = E_{Mol+P} - E_{Mol} - E_P$, where $E_{Mol}$, $E_P$, and $E_{Mol+P}$ are the energies of the molecule, phosphorene sheet, and molecule adsorbed phosphorene, respectively.

## 3. RESULTS AND DISCUSSION

**3.1. Electronic properties of phosphorene under compressive strain.** For the supercell initially in the planar configuration, a period of ripple is formed upon applying a period of sinusoidal out-of-plane displacements to the atoms along the zigzag (or armchair) direction, followed by an energy minimization by using VASP. The length of the ripple corresponds to the wavelength of the sinusoidally shaped phosphorene along the zigzag (or armchair) direction. Since periodical boundary condition with a fixed period is applied laterally, that is, along the armchair (or zigzag) direction, the lattice spacing along the armchair (zigzag) direction remains unchanged.

Fig. 1a-e presents the relaxed rippled structure (upper panel) and the variation of band gap (lower panel) of monolayer phosphorene under compressive strain of 0%, -5%, -20%, -25% and -35%, respectively, along armchair direction. It is seen that the phosphorene remains a direct band gap semiconductor within the strain range. For the compressive strain in the range from 0% to -20%, the band gap decreases from 0.84 eV to 0.51 eV. Our calculated results for the strain range from 0% to -10% are in a good agreement with previous results[19]. For the compressive strain higher than -20%, the band gap achieves approximately a constant value at about 0.51-0.54 eV.

Fig. 1f-j presents the relaxed rippled structure (upper panel) and the variation of band gap (lower panel) of monolayer phosphorene under compressive strain of 0%, -13%, -20%, -25% and -30%, respectively, along zigzag direction. It is seen that the band gap decreases rapidly from 0.91 eV to 0 eV as the compressive strain increases from 0% to -30%. Meanwhile, the conduction band minimum (CBM) shifts from $\Gamma$ to a point between the $\Gamma$ and Y points (see Fig. 1h and i), indicating a direct-to-indirect band gap transition. Moreover, at -30% compressive strain and higher, the band gap disappears completely (see Fig. 1j), signifying a semiconductor-to-metal transition. Hence, it is possible to switch phosphorene from a direct band gap semiconductor to an indirect semiconductor and further to a metal simply by applying the compression strain along zigzag direction.

We have also performed hybrid functional (HSE06) calculations to examine the ripple effect. We find that the trends for the bandgap of rippled phosphorene predicted by using both HSE06 and PBE functionals are the same. For example, our PBE calculations predict the bandgaps of ~0.54 eV and ~0 eV for the compressive strains of -20 and -30% along the zigzag direction, respectively, and our HSE06 calculations predict the bandgaps of ~0.86 eV and ~0.3 eV, respectively, under the same conditions. Besides, we also find that both PBE and HSE06 calculations predict similar atomic structures and lattice parameters for pristine and rippled phosphorene. Hence, the predictions by using these two functionals are qualitatively similar. Due to the high computational demand of the hybrid functional (HSE06) calculations, we did not use it to perform calculations for other systems.

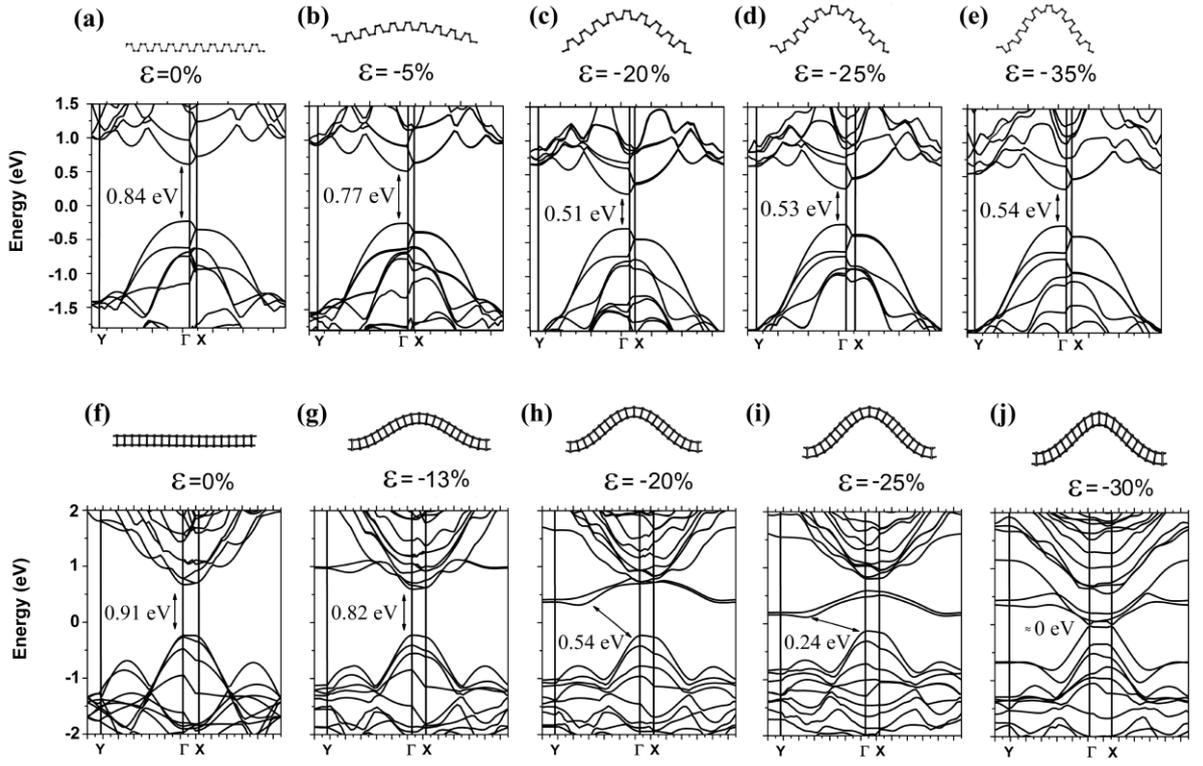

**Fig. 1.** Relaxed structure and variation of band gap of monolayer phosphorene under a compressive strain applied along armchair direction (a)-(e) and zigzag direction (f)-(j). The applied strain is (a) 0%, (b) -5%, (c) -20%, (d) -25% and (e) -35%, (f) 0%, (g) -13%, (h) -20%, (i) -25% and (j) -30%, respectively.

To understand the underlying physical origins for the band gap variation, we investigate the structural deformation of the rippled phosphorene by examining the ripple-induced bonding configuration changes. For both zigzag and armchair directions, we track the variation of bonding configurations of the unit cell on the top peak of the rippled structure as shown in Fig. 2b and c, respectively. The justification for choosing these unit cells is that only the atoms in these unit cells have significant contributions to the VBM and CBM (see below). As defined in Fig. 2a, we track the following parameters during compression deformation: The bond length connecting the hinges $l_1$, the bond length of the hinges $l_2$ and $l_3$, and the hinge angles $\alpha$ and $\gamma$. The calculated relative variations of the bond lengths, that is, $\Delta l_1 = (l_1 - l_1^0)/l_1^0$, $\Delta l_2 = (l_2 - l_2^0)/l_2^0$, $\Delta l_3 = (l_3 - l_3^0)/l_3^0$, and the angle difference ($\alpha - \gamma$) are presented in Fig. 3b and c, respectively.

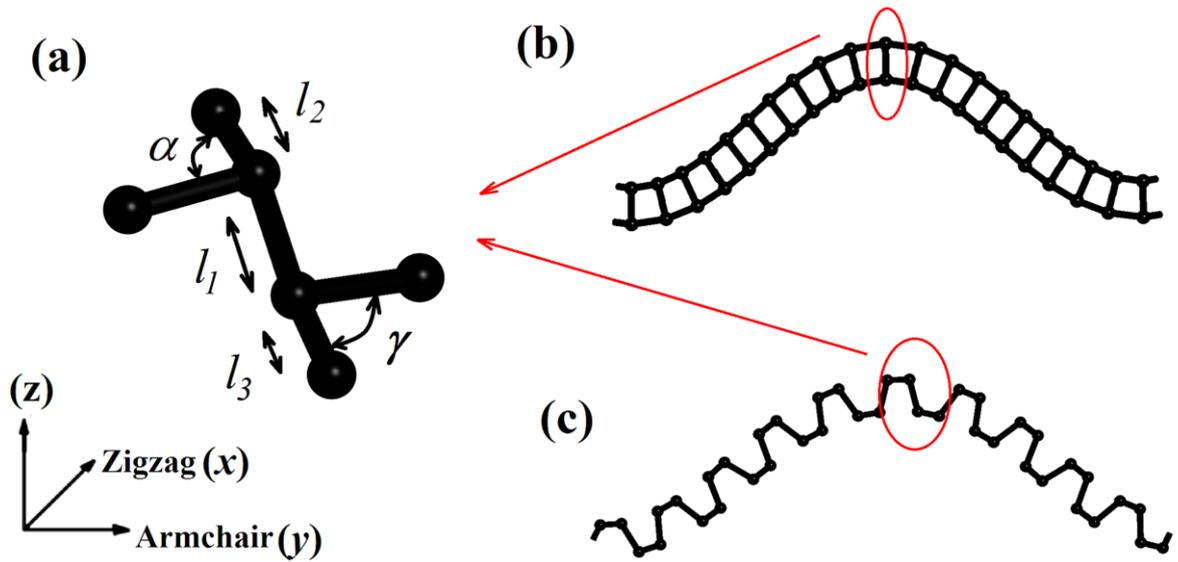

**Fig. 2.** (a) Definitions of the structural parameters for phosphorene. The unit cell at the top peak of the rippled phosphorene is used to analyze the bonding configuration changes under compressive strain along zigzag (b) and armchair (c) direction.

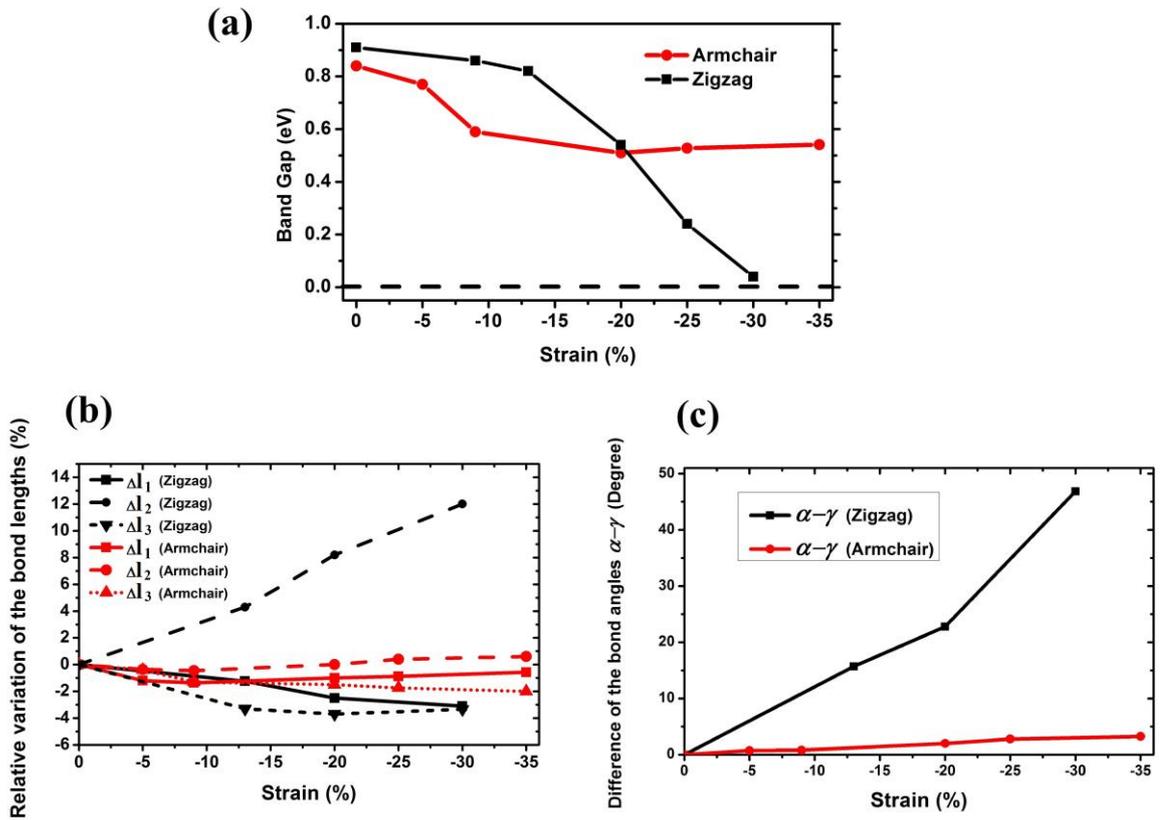

**Fig. 3.** (a) Variation of band gap for rippled phosphorene obtained under the compressive strain along armchair (red line) and zigzag directions (black line). Zero band gap is denoted by the black dashed line. (b) Relative variations of bond lengths $\Delta l_1 = (l_1-l_1^0)/l_1^0$, $\Delta l_2 = (l_2-l_2^0)/l_2^0$, $\Delta l_3 = (l_3-l_3^0)/l_3^0$ and (c) hinge angle difference ($\alpha - \gamma$) for phosphorene under compressive strains along armchair (red line) and zigzag direction (black line).

For the compressive strain along armchair direction, it is seen that with increasing the strain from 0 to -10%, the relative variations of the bond lengths $\Delta l_1$, $\Delta l_2$ and $\Delta l_3$ all decrease, while the angle difference ($\alpha - \gamma$) is roughly unchanged. For the compressive strain from -10% to -20%, the bond lengths $l_1$ and $l_2$ are slightly increased while $l_3$ is slightly decreased, and the angle difference ($\alpha - \gamma$) is slightly increased. For the compressive strain higher than -20%, all the parameters are roughly unchanged. By comparing the behavior of the band gap as shown in Fig. 3a and that of the relative variations of the bond lengths $\Delta l_1$ and $\Delta l_2$ as shown in Fig. 3b, we find the correlation between the band gap change and the bonding configuration changes. More specifically, the band gap decrease is correlated with the increase of the difference between angles ($\alpha - \gamma$), up to -20%; and after that the band gap and the angle difference ($\alpha - \gamma$) are roughly unchanged. Thus, we conclude that the strain sensitivity of the band gap for the compression strain up to -20% comes from the angular distortion of the bonds, which leads to substantial changes in the overlapping of the wave functions of neighboring atoms and thus the variation of the band gap. The small variation of the band gap for the compression strain beyond -20% can be explained by the comparative immutability of the bond angles and lengths.

When the compressive strain is applied along zigzag direction, the angle difference ($\alpha - \gamma$) and the relative variation of bond length $l_2$ are drastically increased while the relative bond lengths $l_1$ and $l_3$ are only slightly decreased as shown in Fig. 3b and c. By comparing the band gap change and the bonding configuration changes, we find that the drastic band gap change is correlated with the changes in the variation of the bond length $\Delta l_2$ and the angle difference ($\alpha - \gamma$). It is these bonding configuration changes that lead to the significant change in the band gap.

To further understand the behavior of the band gap, we also consider the density of states (DOS), the partial density of states (PDOS) and the local density of states (LDOS) of rippled phosphorene under the compressive strain applied along both armchair and zigzag directions. Figure 4 shows the PDOS of phosphorene under the typical strains of 0, -20% and -35% along armchair direction and 0, -20%, -30% along zigzag direction, respectively. For the uncompressed phosphorene, that is, under zero strain, the $s$, $p_x$, $p_y$, $p_z$ states are nearly equally distributed across the whole energy spectrum except within the band gap, where the frontier state is only comprised of $p_z$ orbital in both cases, which is consistent with previous work[53].

As shown in Fig. 4b, at -20% strain along armchair direction, the contribution of $p_y$ state to both the valence band maximum (VBM) and the conduction band minimum (CBM) increases and the VBM shifts towards the CBM. Under -35% strain as shown in Figure 4c, the gap between the VBM and the CBM is nearly unchanged and $p_y$ and $p_z$ states have an equal contribution. Therefore, the contribution of $p_y$ state to the VBM and the CBM increases accompanying the change of the band gap. As shown in Fig. 4e and f, at -20% and -30% strains applied along zigzag direction, the CBM shifts towards the VBM and the contribution of $p_x$ state is increased with the increase of the compressive strain. There is an important difference between the flat phosphorene and rippled phosphorene: For the flat phosphorene, in the valence top, only $p_y$ state is present while $p_x$ state is clearly absent, signifying the linear dichroism in flat phosphorene[53]. For rippled phosphorene along the zigzag direction, a strong mixing of $p_x$ and $p_y$ states occur in the valence top, suggesting the disappearance of linear dichroism in the rippled phosphorene.

We next examine the localized electronic properties by looking into the PDOS of atoms on different parts of the ripple. We are particularly interested in those atoms that are highly strained and located at the peaks. Fig. 5a and c shows the rippled phosphorene configurations at strain of -20% along armchair and zigzag direction, respectively. PDOS of the atoms marked by "1"-"4" are considered. For the rippled phosphorene compressed along armchair direction, the main contribution to the CBM is from $p_z$ states of atoms "1"-"4", with dominant contribution from atom "1", as shown in Fig. 5b. In contrast, for the zigzag case, as shown in Fig. 5d, the main contribution to the CBM is from $p_x$ states of atoms "1" and "3", which experiences the largest tensile strain. We can see that these atoms make the largest contribution to the states from 0.1 to 0.5 eV, which are related with the flat bands around the Fermi level in Fig. 1h. This means that the flat bands are related to the localized states of these atoms, which become less bonded and passivated with their neighbors. Consequently, these atoms, which are on the top (atom "1" and "3") and under high tension, are responsible for the reduction of the band gap.

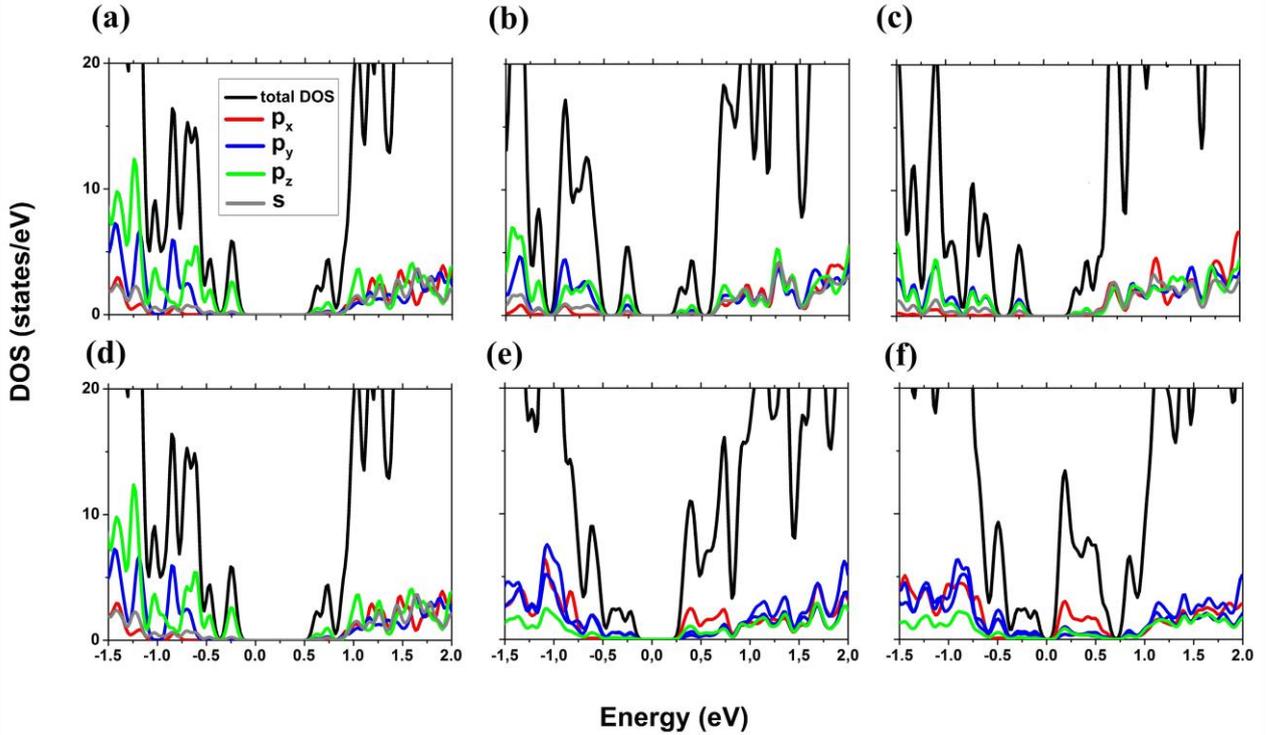

**Fig. 4.** DOS and PDOS of the rippled phosphorene under the compressive strain: 0% (a), -20% (b), -35% (c) along armchair direction, and 0% (d), -20% (e), -30% (f) along zigzag direction, respectively. For each case, the electronic

states *s*, $p_x$, $p_y$, $p_z$ are represented by grey, red, blue and green lines, respectively. The total DOS is plotted as the black line.

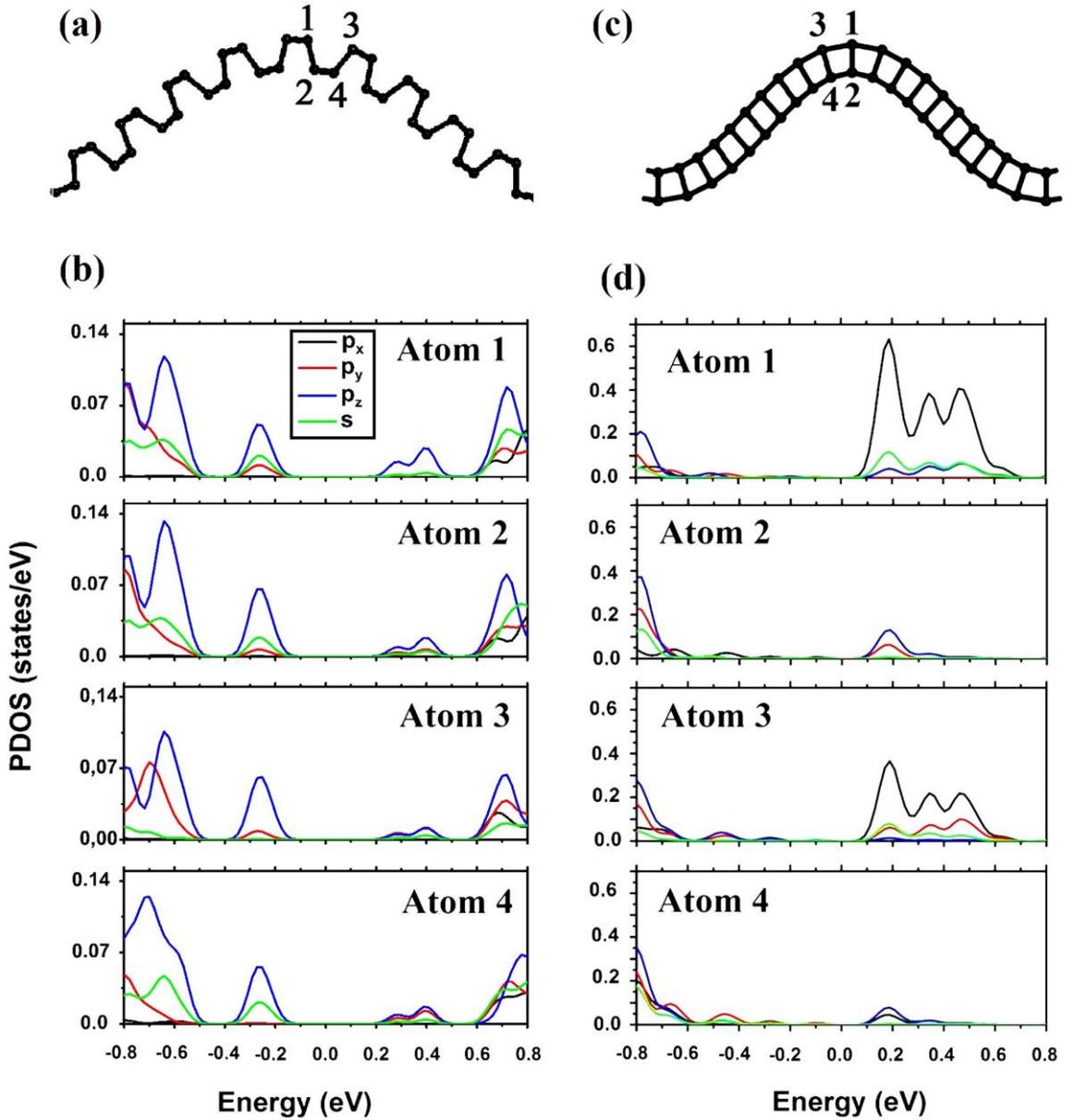

**Fig. 5.** Rippled phosphorene under a compressive strain of -20% along (a) armchair and (c) zigzag direction, respectively. (b) and (d) show the partial density of states (PDOS) for the atoms "1"-"4" in (a) and (c), respectively.

To present the spatial variation of the electronic structures, we plot the LDOS of each pair of atoms along the line profiles of the ripples in Fig. 6a and b under compressive strain of -35% along armchair and -30% along zigzag direction, respectively. For the ripples along armchair direction, it is seen that the local gaps of each pair of atoms are nearly overlapped, and there is no shift in the LDOS curves. On the contrary, for the rippled phosphorene along zigzag direction, the LDOS of each atom pair shifts relatively towards each other. A red-shift of the states is predicted for atoms with positions approaching the peaks, while a blue-shift of states for atoms located in the middle part of the wave-like profile. This explains the variation of band gap and confirms the extraordinary tunable properties of rippled phosphorene. Such spatially dependent alignment of valence and conduction bands, which is similar to that in the type-II semiconducting heterostructures, is able to facilitate the separation of electrons and holes to different parts. Moreover, this spatial modulation of the band gap is also able to induce the funnel effect in a reduced spatial region. Funneling has been recently proposed as a powerful strategy to enhance the efficiency of photovoltaic energy harvesting devices by facilitating the collection of photogenerated carriers[54,55]. Our results suggest that the electronic structure of rippled phosphorene formed under compressive strain along armchair direction is more robust than that along zigzag direction. The latter should present strong spatially-dependent behavior of the local quantities, such as electronic and optical properties, which allow the engineering of rippled phosphorene by taking advantage of its extraordinary flexibility.

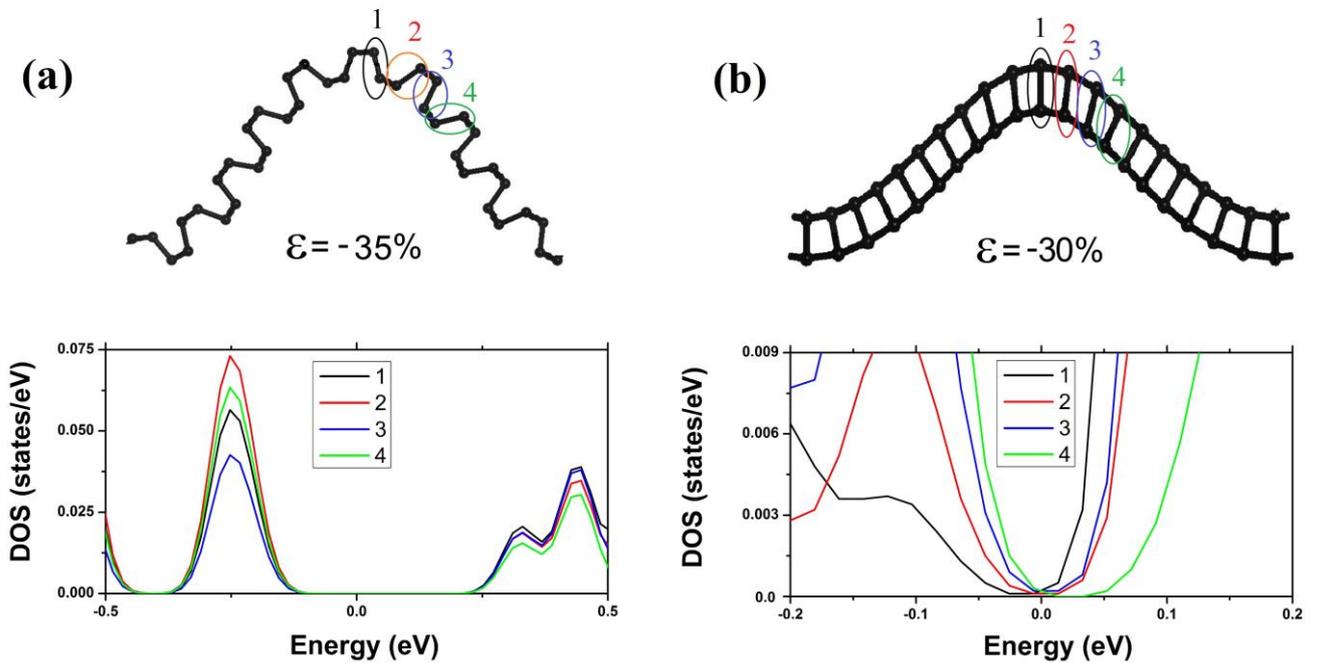

**Fig. 6.** The LDOS for the atom pairs of rippled phosphorene compressed along (a) armchair direction with a strain of -35% and (b) zigzag direction at a strain of -30%, respectively. The atom pairs are marked as "1"-"4" and the color of marker corresponds to the color of the line for the LDOS.

**3.2. Physisorbtion of NO gas molecule on rippled phosphorene.** Atomically thin 2D materials are highly flexible and tend to form ripples or wrinkles in either suspended or supported sheets[54,56]. Curvatures due to ripples are able to induce inhomogeneous deformations of the lattice and modify the chemical properties, as shown in graphene and CNT[41-47]. Below, we consider the influence of ripples on the chemical activity of phosphorene. Absorption energy of NO gas molecule on both planar and rippled phosphorene, and the charge transfer between the molecule and planar/rippled phosphorene are investigated. We have examined several possible positions on the highly symmetric sites of NO molecule on the both planar and rippled phosphorene, including both above the puckered hexagon and the zigzag trough, with the molecules being aligned either perpendicular or parallel to the surface. To examine the curvature effect on the adsorption, the NO molecule adsorbed at both the concave and convex regions of the ripple are considered. The lowest-energy configurations of the NO molecule adsorbed at the concave and convex regions are shown in Fig. 7a and b, respectively.

It is found that for both cases, the absorption energy $E_a$ decreases with increasing the compressive strain for the position both above and below the phosphorene as shown in Fig. 8. The amount of charge transfer between the NO and phosphorene is obtained by using the Bader analysis[57]. For the planar case, $E_a$= -0.29 eV, consistent with previous studies[25,29]. We find that NO gas molecule accepts electron upon adsorption on phosphorene and the total amount of transferred charge from phosphorene is 0.084 $e$. With increasing the strain from 0 to -30%, the acceptor ability of NO molecule increases and the total amount of transferred charge increases up to 0.206 $e$. Such changes can be attributed to the curvature effect, which modifies the local carrier density and orbital hybridization as shown in the previous section. Therefore, ripple-induced deformation is an effective way to promote the chemical activity of phosphorene in terms of absorption energy and charge transfer, and rippled phosphorene may be useful for gas sensing applications and for improving the doping efficiency by adsorbing chemical species.

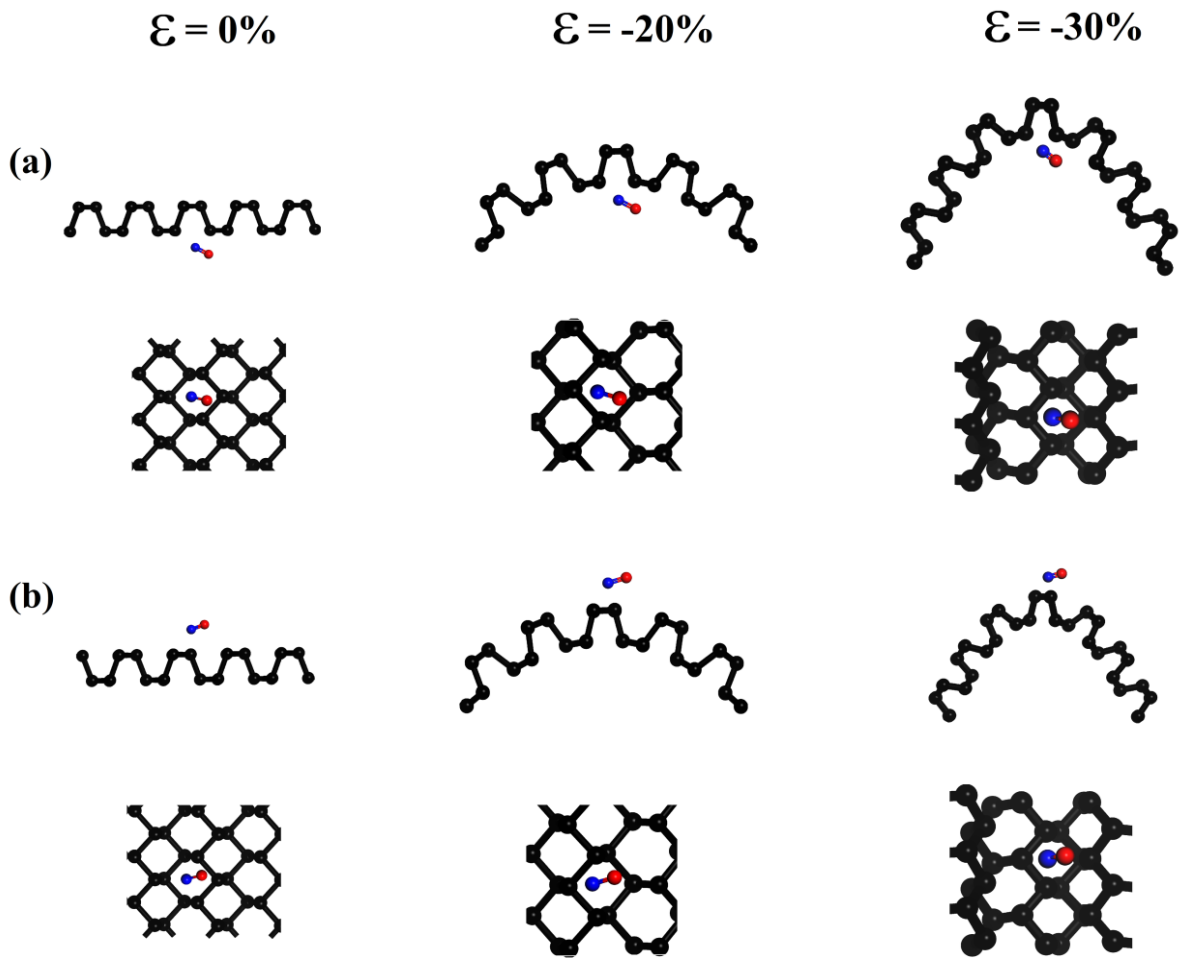

**Fig. 7.** The most stable adsorption positions of NO gas molecule (a) below and (b) above planar/rippled phosphorene surface. The balls in black, blue, and red colors represent the phosphorus, nitrogen and oxygen atoms, respectively.

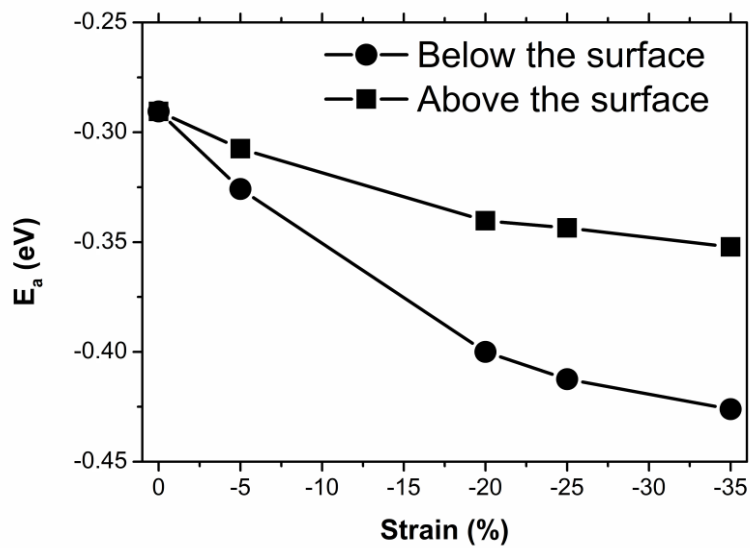

**Fig. 8.** Variation of the absorption energy of NO gas molecule on phosphorene surface with increasing the curvature by applying compressive strain.

## 4. CONCLUSIONS

By using first-principles calculations, we reveal that ripples can lead to significant changes in the electronic properties of phosphorene. The strong spatial dependence of the electronic structure in rippled phosphorene along the periodic line profile may potentially allow the control of the transport of carriers via ripple engineering. Our work may explain the recent experiment that observes the spatially dependent optical properties in rippled phosphorene, where periodic ripples with large curvatures were obtained by transferring phosphorene to a greatly pre-stretched elastomeric substrate, followed by a relaxation of the pre-strain in the substrate[28]. In addition, we also show that rippled phosphorene is able to promote the adsorption of NO molecule and increase the charge transfer, signifying an enhanced chemical activity. The marked enhancement in the chemical activity suggests that rippled phosphorene is a promising material for gas sensing applications.

## AUTHOR INFORMATION

**Corresponding Authors**

*E-mail: caiy@ihpc.a-star.edu.sg.

*E-mail: kzhou@ntu.edu.sg.

## ACKNOWLEDGMENTS

The authors gratefully acknowledge the financial support from the Ministry of Education, Singapore (Academic Research Fund TIER 1 - RG128/14), the Agency for Science, Technology and Research (A*STAR), Singapore and the use of computing resources at the A*STAR Computational Resource Centre, Singapore. Sergey V. Dmitriev acknowledges financial support from the Russian Science Foundation grant N 14-13-00982.

**TOC:**

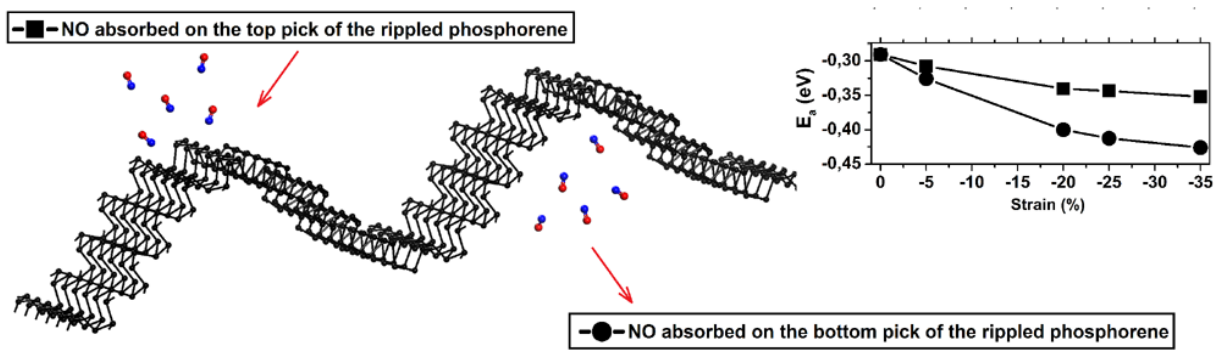